# A World of Views:
## A World of Interacting Post-human Intelligences

By Viktoras Veitas and David Weinbaum (Weaver)[†]

**Part I The long tail of the Singularity[1]**

**1 1001 times more intelligent?**

What would a human hundreds or thousands times more intelligent than the brightest human ever born be like? We must admit we can hardly guess. A human being of such intelligence will be so radically different from us that it can hardly, if at all, be recognized as human. If we had to go back along the evolutionary tree to identify a creature 1000 times less intelligent than the average contemporary human, we will have to go really far back. Would it be a kind of a lizard? An insect perhaps? Considering this, how can we possibly aspire to have a grasp of something a thousand times more intelligent than us? When it comes to intelligence, even the very attempt to quantify it is highly misleading. Now if we attend to a seemingly adjacent question, what would a machine with such capacity for intelligence be like? Just coming up with an approximate metaphor requires a huge stretch of the imagination, meaning that almost anything goes… What would a society of such super intelligent agents, be they human, machines or an amalgam of both, be like? Well, here we are transported into the realm of pure speculation. Technological Singularity is referred to as the event of artificial intelligence surpassing the intelligence of humans and shortly after augmenting itself far beyond that. It is no wonder that the mathematical concept of singularity has become the symbol of an event so disruptive and so far reaching that it is impossible to conceptually or even metaphorically grasp, much less to predict.

We do not know whether a Singularity will ever happen or what it will be like. Yet, the disruptive events already taking place in our lifetime at an ever increasing rate clearly indicate that a great something else awaits us all in a future too near to comfortably disregard. It indeed may be worth, being the conscious observers that we are, to relate to our situation as on the brink of Singularity. We think that this brink of Singularity situation is more important and informative than any speculation about the Singularity itself. Therefore we focus here on understanding *the process of sociotechnological evolution* which seemingly takes us towards a Singularity, whatever that may mean. Moreover, we think that this process may be the most significant evolutionary transition that ever took place in the history of humanity and probably in the history of life on this planet. How does this process work? What consequences may it bring to us as individuals and as a society? Can we somehow influence it, direct it? Can we make it sustainable and graceful, introduce some sense of continuity into an overwhelming experience of discontinuity? The image of a World of Views - a world of diversity on the brink of Singularity - will shed some light, we hope, on how to start answering these questions.

---

[†] Both authors contributed equally to this work.
[1] The title refers to the historical process of accelerating change that did not start recently but rather began at the dawn of humanity as a tool making civilization.

**2 Escaping the constraints of natural evolution**

The Latin word *socium* means companion, associate. Likewise, by *sociotechnological* we mean living with technology, disposed to symbiotic relations. We describe society as a complex, highly connected, network of agents which possesses, as a whole, organic, life-like characteristics such as self-organization, self-regulation, coordination, adaptive behavior and more. We refer to this network as the *sociotechnological system*, and to its dynamics as the *sociotechnological evolution*. We do not associate agents just with humans; we rather take them to be any entity which can be assigned with a degree of autonomy, intentionality and identity [21]. For now, these are mostly humans mildly augmented by technology[2]. On the brink of Singularity, these may be heavily augmented humans (i.e. cyborgs) or technological artifacts mildly augmented by humans (e.g. self-driving cars), not to mention autonomous organizations involving both humans and machines (e.g. corporates, autonomous management systems, cities). The point is that the specific characteristics of future agents do not matter. Society as a super-organism is mostly shaped by the interactions among all the agents in the network, and much less by the particular characteristics of individual agents.

We replace the vague term of Singularity with an understanding of the sociotechnological evolution as the process of intelligence escaping the constraints of biology. It is a process with a very long history which actually started with humanity arising as a tool making civilization about 2.5 million years ago. While we may feel that the brink of Singularity is special to our times of accelerating change, a somewhat similar experience may well had been shared by witnesses of the Scientific Revolution of the 17th century, or those frequenting the library of Alexandria.

Intelligence escaping the constraints of biology is evidently a process of a continuous transition rather than a singular event. This is the process where humanity as a biologically evolved form of intelligence undergoes a transition from its biological stage into a post biological stage. A post biological stage means primarily the blurring of distinctions between the natural and the artificial. It is characterized by the ever increasing prominence of non-biological forms of embodiment and agency (e.g. anything from bionic limbs and artificial organs, to semiautonomous avatars, highly autonomous robotic systems and semi sentient computing agents). This however does not necessarily pronounce the eventual disappearance of biological structures. It rather means that future structures of biological origin will be shaped by technology more significantly than by natural evolution (e.g. genetic manipulation, but also other interventions at both the cellular and organism levels). In this process of sociotechnological evolution, humanity undergoes an *intelligence expansion*[3] via the mediation of technology.

---

[2] Cyborg Luddite Steve Mann on Singularity 1 on 1: Technology That Masters Nature is Not Sustainable. http://www.singularityweblog.com/cyborg-steve-mann/

It is difficult to turn Singularity into a verb, but we propose and even urge to think about the brink of Singularity in terms of a process. Being complex and unpredictable, this process will most probably not yield the scenarios that we project, whether utopian or dystopian. Nevertheless, it does have a direction that we can better characterize and understand.

## 3 Sociotechnological evolution and intelligence expansion

What is sociotechnological evolution and how does it drive intelligence expansion? Intelligence is quite difficult to formally define but is most apparent in the behavior of agents when observing their interactions with their environment. An intelligent behavior is achieved via a progressive process of developing models of the environment and acting upon them. The better an agent is at developing models, the more capable it is of anticipating the dynamics of its environment and responding appropriately.

Forming effective model representations of the environment is associated, on the one hand with general intelligence and on the other hand with evolutionary fitness. Natural evolution persistently selects those organisms that operate the best models. This means that in a profound manner, evolution selects for intelligence. Moreover, the evolution of nervous systems and brains can be understood as strategies to achieve more accurate and faster adapting models of complex, fast changing environments compared to what mere genetic modification can achieve. We suggest that sociotechnological evolution is the continuation of this biological trend, only augmented by technology.

Technology greatly extends the capability of humans to create better models of their actual environment. It also provides them with the tools to effectively project desirable, not yet actualized, states of their environment.

---

[3] Contrary to I.J. Good's concept of intelligence explosion' [12], 'expansion' embraces all variants of increasing intelligence: both punctuated ultra quick 'explosions' and gradual developments. We emphasize the significance of the overall underlying process.

Finally, it enables them to modify their environment to fit these projections. This is, in essence, how tool making (language and scientific thinking included) initiated the powerful positive feedback mechanism that drives sociotechnological evolution. Increase in capacity to model the environment actually amplifies the capacity to modify the environment because the availability of reliable representations allows for more effective and directed interventions. A changing environment, in turn, exerts an adaptive pressure to increase the capacity of modeling and necessitates knowledge and intelligence expansion. These effects are mutually reinforcing. Consider for example the agricultural revolution; agriculture started with very basic knowledge of how to cultivate plants and animals for food. This knowledge enabled humans to settle and then modify their environment in order to produce a reliable supply of food instead of being dependent on hunting and gathering which were much less predictable practices. Controlling the production of food brought, what was probably considered then, an age of abundance, but it also greatly changed the way humans lived. Settlements grew, became more complex to manage and a variety of new challenges arose, e.g. production of tools, storage of goods, forecasting weather, building shelters, managing waste, etc. Solving these problems stimulated the expansion of knowledge and specialization. The expansion of knowledge, in turn, has placed in the hands of humans more means to further modify and expand their settlements, which gave rise to further problems such as division of labor, exchange of goods and knowledge among strangers, etc. The development of human settlements was coupled with the expansion of knowledge and technological means. These were indeed mutually reinforcing processes. A combination of better nutrition, the need to navigate in more complex social situations (and perhaps more free time to think) which made humans smarter and more creative in general, further amplified knowledge expansion. This positive feedback mechanism operates since then at different scales beginning with single agents, their communities and organizations, and extending to the largest scale of the sociotechnological system and the whole planet[4].

    The coupling of the sociotechnological system with its environment is highly complex and exhibits a variety of nonlinear effects. The result of this is that the system becomes less stable and much less predictable, which is well illustrated by the behavior of financial markets and the global economy [31]. It seems that presently, while the overall acceleration has the advantageous effect of intelligence expansion, the capability to adequately model a fast changing complex environment lags behind the actual dynamics of change. This has many adverse effects: actions based on unreliable models result in destabilization,

---

[4] A Brief History of Intellectual Discussion of Accelerating Change. http://www.accelerationwatch.com/history_brief.html

unpredictability and dangerous if not destructive policies. Such effects may indeed catalyze the reorganization of the sociotechnological system into its new evolutionary phase, but they may also be catastrophic. Which kinds of instabilities are probable to develop largely depend on the organization of the sociotechnological system and its adaptive capacities. Based on this cybernetic perspective of the sociotechnological system and its dynamics, we propose our vision of a *World of Views*. With this vision we hope to outline a sociotechnological organization that can successfully cope with the challenges of the accelerating transition humanity is undergoing.

**4 A World of Views**

*A World of Views is a vision of a future sociotechnological system that embodies a multiplicity of unique, modular and open co-evolving worldviews.* What is a worldview? A worldview is both a formal philosophical concept and a term of everyday use that comes to describe a unique perspective of an agent at various levels from individuals to organizations, communities and whole societies. A worldview can be understood as a gestalt perception, both individual and collective, in relation to self, others, society, and the cosmos at large. A worldview may exist on many levels and contain contradictions and paradoxes. Every worldview is selective, not only in regards to which categories of the agent's attributes are included, but also in relation to the facts which are asserted to be true of these attributes. Some views are narrow, ignoring many possibilities, others are more comprehensive. Each, however, selects which attributes and qualities are to be considered real, which are to be developed, admired, accepted, despised or otherwise attended to. These views are held at varying degrees of awareness by individual agents as well as by collectives of agents (e.g. communities, organizations, societies) [25]. Since the concept is applied with different meanings in different contexts, it is useful to distinguish three planes that together constitute a worldview: the systemic/objective, the subjective, and the intersubjective. The first plane consists of given facts (e.g. the cyclic change of seasons) and their organization. This plane is mostly shaped by the sciences. The second plane consists of individual subjective experiences and the belief systems they induce (e.g. this person is trustworthy). The third plane consists of consensual structures that arise in the social interactions among humans (e.g. all men are born equal). Worldviews often tend to emphasize the importance of one of the planes and understate the role of the others. But it is the interplay among the three planes that accounts for the complexity and nuanced structure of worldviews. A comprehensive discussion of the concept and a wealth of further references can be found in [35, 36].

In our vision, besides the inherent uniqueness that characterizes a worldview, well-structured worldviews must have two additional characteristics:

*modularity* and *openness*. Modularity means that worldviews are assembled from modules where each module is an assemblage of ideas, beliefs, values, representations and operational strategies that constitute together a coherent and relatively independent whole. The idea is that worldviews can develop and evolve by adding, omitting, or replacing modules. Openness is the capability of a worldview to accommodate and interact with other worldviews. An open worldview is adaptive and interactive with other worldviews, even those with rival perspectives, beliefs or values. Worldviews which are both modular and open can share, exchange and co-evolve.

From a societal perspective, a World of Views is a heterogeneous network of agents such as humans, technological artifacts with diverse degrees of autonomy, intelligence and expertise, organizations combining both humans and artifacts and possibly other futuristic agencies (e.g. augmented apes, dolphins, cyborgian entities etc.). These diverse agents pursue a life style befitting the unique worldview they hold. Worldviews can of course partially overlap or be in conflict with each other. We envision a fluid social order facilitated by *distributed social governance* (see section 9) which is very different from the social order we live by today. Distributed social governance is not based on fixed hierarchical systems of governance such as nation states or supranational institutions. It is rather based on principles similar to open source projects, art biennials or academic collaborations: the sharing of ideas, perspectives, values and goals and the formation of fluid *ad hoc* social coalitions according to emergent relevance or interest. Yet, in a World of Views, co-evolution and sharing is not a requirement but rather an option for agents that may wish to engage and cooperate.

As we will further discuss, one of the most important consequences of the future will be *abundance* [9, 10]: the elimination, in the practical sense, of constraints that force agents to compete over limited resources. A world of abundance will be a world of much fewer existential constraints and many more existential options, without fear. Therefore, one world as a methodological construct[11] will become a choice of the inhabitants of the world rather than a constraint of sociotechnological evolution. In a World of Views, survival is no longer an ultimate driving force but rather a taken-for-granted baseline. The shift in the very forces that shape the sociotechnological evolutionary process is one of the most profound social changes we foresee. On the brink of Singularity, where the physical constraints that were and still are guiding biological and social evolution are gradually relaxed, it is the images, narratives and values constituting our worldviews that will take their place in shaping the evolutionary process. Our vision is a vision of radical pluralism and openness that catalyzes and is being catalyzed by sociotechnological evolution. A World of Views is a world that thrives and evolves thanks to the enormous diversity and variation of intelligences, their chosen embodiments, styles of expression and co-evolution in the form of multiple overlapping and fluid social institutions. In the following sections we develop the rationale behind this vision and argue not only the fact that it is a realistic outcome of sociotechnological evolution, but also why we think it is a desirable outcome.

**Part II Bootstrapping our future**

**5 Framing the forces at play**

Nietzsche, the post-modern philosopher, based much of his philosophy on a view of the world as a complex system of interacting forces (known as "will", in his terminology). In his typology of forces he distinguished between reactive forces and active forces. Reactive forces are always forces that react to given situations, they respond and adapt but will never initiate change. They lack the spontaneity and creativity associated with active forces. Active forces are spontaneous and creative - the real movers and catalysts of change according to Nietzsche. They are expressions of freedom as they are never bound to a given state of affairs or a prescribed order [8]. In this part, we will use this idea to highlight the profound change that we foresee in the future dynamics of the sociotechnological system.

As individuals we move and act driven by two kinds of motives that are analogous to the two kinds of wills mentioned by Nietzsche: either we find our current situation dangerous or deficient in some significant aspect (i.e. we react), or, we are pretty happy with the current situation but nevertheless desire something else, better, more interesting and more fulfilling. The first kind of motives deals with deficiencies while the second has to do with the need for growth and self-actualization [26]. In other words, we suggest here a distinction between change driven by constraints and change as an expression of freedom. If we relate to the sociotechnological system as a higher order cybernetic organism, the same distinct kinds of forces can be associated, on a different scale, with societal change at large.

In the following sections we develop arguments showing that *both kinds of forces* that are driving change in the sociotechnological system are driving it towards the same direction: the realization of the future society as a World of Views. We show how our vision of a World of Views is a desirable *response* to the critical problem of coping with the instability and unpredictability that are inherent in the current phase of accelerating change. Given that the sociotechnological system will avoid the existential risks inherent in its very dynamics, we can expect a future of abundance. A sociotechnological evolution under circumstances of abundance will, with high probably, culminate in realizing our vision. In other words, a World of Views will emerge as a manifestation of *active forces* in a state of abundance. We reason therefore that a diversified, open, co-evolving social dynamics is both catalyzing and being catalyzed by sociotechnological evolution. In that, we identify a plausible positive feedback mechanism that actually bootstraps our vision.

**6 Our fragile present**

It seems paradoxical that our ever increasing knowledge of the world and technological progress make us more fragile. Where does this fragility come from and how can we cope with it? We think that understanding the big picture of sociotechnological evolution provides important clues.

**6.1 Understanding fragility**

In his recent book Antifragile [32] Nassim Taleb gives the following definition to fragility:

> "[..] what does not like volatility does not like randomness, uncertainty, disorder, errors, stressors, etc. Think of anything fragile, say, objects in your living room such as the glass frame, the television set, or, even better, the china in the cupboards. If you label them "fragile," then you necessarily want them to be left alone in peace, quiet, order, and predictability. "

With cybernetic systems, that is, systems that involve feedback mechanisms, fragility is much more complicated because such systems cannot be left in peace and fragility therefore is not a static characteristic of an object but rather a dynamic property of a system. Anyone who has ever suffered a painful injury is well familiar with how resourceful one becomes in avoiding any movement or change of posture of the painful body part so much, that it becomes entirely rigid while other healthy parts compensate and adapt accordingly. There is much sense in such response: pain signals vulnerability and avoiding it means allowing the injured area to rest and recover quicker. Yet, often, one discovers that even after the injury is long healed and there is no pain, there is a residue of physical rigidity left and even some instinctive fear associated with using the once injured body part again. It is as if our body has forgotten its healthy normal condition and now prefers a safer though limited condition. But the effect of fragility does not end here. The self-limiting protective behavior originating from the body's attempt to (over) protect a fragile point may, in the long run, cause it to lose its normal range of variable response and leave it eventually more vulnerable than before the injury. Medical practitioners are aware of this phenomenon and help patients, sometimes against their own judgment, to return as quickly as possible to regular patterns of activity inasmuch as this is possible.

The important lesson here is that under certain conditions, fragility begets more fragility, increasing the probability of major failures due to random unpredictable shocks. Every system, no matter how robust, has points of vulnerability. But what is important in the context of cybernetic systems and perhaps critical in the case of the sociotechnological system is not so much identifying such points but rather focusing on understanding the mechanisms that amplify fragility and learning how to modify them. It would not be an overstatement to say that mechanisms that amplify fragility constitute an existential risk to humanity as utterly unpredictable minor shocks become implicit seeds to system-wide disastrous events. This is especially true taking into account the potentially ruinous effects of the system, greatly empowering the capacities of individuals: taking advantage of knowledge available to all and with relatively humble financial means, a person can build a nuclear device, develop a lethal biological agent, or deploy a viciously damaging computer virus.

This is, in a nutshell, modern day fragility.

**6.2 The consequences of a complex world**

Our world becomes progressively more fragile because the current sociotechnological dynamics has a growing number of powerful amplification processes. These processes placed vast powers in the hands of humans, but by the same token they also amplify the effects of errors and accidents. It all comes down to technology because technology in its deepest sense is all about amplification - doing more with less. Cars, ships and planes enable to travel farther and faster. Writing (and electronic media) enables to remember more data for longer. Communication enables the distribution of more information farther and faster. Machines and factories enable the production of more of everything including more machines, and computers enable more knowing, predicting, modeling and inventing. In this process, computing amplifies technological development and is the primary catalyst of technology-induced change.

The downside of amplification is not difficult to understand. Let us examine a number of more concrete factors, all of which are associated with the sociotechnological system becoming a very complex system (for an excellent treatment of the subject see [15]).

> **Hyper-connectivity** is a major symptom of progress, resulting in our world becoming progressively more connected at many levels: starting from global transportation networks that carry goods as well as people across the planet and culminating in global communication networks, that mobilize and distribute information at an ever increasing volume and speed. A hyper-connected world is a world where every agent is connected by numerous means to many other agents and where distances, both in space and time, are collapsing [39]. The effect of collapsing distances is that local perturbations may swiftly spread and become global perturbations. This means that highly unpredictable and initially unnoticed local events may yield global impacts within very short time spans. Local outbreaks of diseases are hardly contained and may spread via the global transportation system and become pandemics. Computer viruses can spread and disrupt vast portions of the Internet with estimated damages of many millions of dollars [19]. A thoughtless rumor can initiate a chain reaction that may crash international financial markets. These are only a few examples of how hyper-connectivity amplifies fragility. It is almost inconceivable for the global system to protect itself from all these local, potentially disruptive perturbations. Nevertheless, like in the analogy of the injured body in subsection 6.1, the tendency of the current sociotechnological system will be to limit the freedom of its components and to guide them into more predictable patterns of behavior. Such systemic response will only increase its fragility in the long term [3].

**Reflexivity** is a concept referring to circular relationships between cause and effect as each element is both affecting and is being affected by other elements. Especially it refers to a feedback relationship between observer (i.e. intelligent agent) and observed (i.e. environment): any examination and action of agents bends the environment and affects the perception and further decisions by the same agents. From its very definition, the sociotechnological system is a reflexive system with a vast number of feedback loops. Reflexivity, blurring the distinction between causes and effects, makes systems difficult to analyze and predict. The contribution of reflexivity to the fragility of the sociotechnological system depends on the kind of feedback mechanisms that operate. A negative feedback has a stabilizing effect on the system's behavior as it resists any change in the state of the system. This is not the case with positive feedback that has the opposite effect of destabilizing the system by amplifying any disturbance. The crucial aspect of the reflexivity property for the sociotechnological system is that patterns of modeling and representation of the world (see 4) have a decisive effect on the type of feedback loops which develop in it. Consider, for example, a stock market crash caused by a positive feedback: a price of a stock randomly fluctuates down which may bring stressed traders to sell that stock because they predict a further decrease. Which indeed becomes a self-fulfilling prophecy: each sale order further reduces the price and drives an avalanche of sale orders which may eventually crash the stock price [31, 3]. In all similar cases, reflexivity is clearly a potential amplifier of fragility.

**Accelerating change** is a positive feedback effect influencing the speed of the overall system's dynamics. As more information is fed to agents they produce more events in response, which produces in turn even more information for other agents. However, the overall capacity to process information relevantly is limited. Accelerated change amplifies fragility because at the level of the sociotechnological system, once the pace of events exceeds the capacity of the system to process information, the system becomes blind and more exposed to unpredictable adverse events.

Understanding the mixed blessing of technology as a systemic amplifier is only part of a deeper puzzle. Arguably we could compare human technology to biology as *a different kind of technology*. Also biological technology is all about amplification: the species that succeed to make more of itself with the least available resources is the species that manages to survive. Yet, life at large is anything but fragile. This profound difference tells us that amplification is perhaps important to understanding fragility, but it is not the whole story.

**6.3 The Clockwork that never was...**

The critical factor of sociotechnological fragility is the manner humans approach technology. This approach seems to be rooted in what we call the Newtonian worldview [18]. This worldview, originating in Newton's days, is a perception of the universe as a vast deterministic clockwork whose components from the minutest to the largest operate in tandem with ultimate accuracy and efficiency. The scientific revolution of the 17th century has established the belief that it is within the powers of the scientific method to fully understand this mechanical universe, its laws and how to control it. Though this perception was quite shaken since the beginning of the 20th century and on (e.g. by quantum theory with uncertainty, chaos theory with unpredictability, and complexity science with uncontrollability), it is still prevalent and reflected in the way we create models of the world, structure organizations and design technological artifacts. From governance systems, armies, supply chains and air traffic control, to the miniaturized electronic chips, all are highly accurate and interdependent machines optimized for efficiency.

As systems become larger and more interconnected, tight interdependence means fragility. The failure of one component may cause, with increasing probability, the failure of the whole system. According to the mechanistic Newtonian view, failure is not natural and is not expected. Of course error margins and failure modes are part of engineering, but systems are not designed to fail (and recover). Hence, the larger the mechanism, the more fragile it becomes because there are so many more components that are not supposed to fail but eventually do. Biological technology did not evolve that way. Biological systems continuously fail and recover as components compensate each other.

Every day an average person loses thousands of neurons in his brain[5] with little effect on mental performance. In comparison, the failure of a single bit on a single silicon chip may cause a major malfunction of a large computer system with thousands of chips. Human technological artifacts indeed work and amazingly so, but because of the way they are designed, when they fail they fail spectacularly as well.

The problem with the deterministic worldview becomes much more acute when it comes to constructing models of the complex fast and changing sociotechnological system. In a world which was fairly stable, less hyper-connected, reflexive and accelerating compared to ours, the basic assumptions of the deterministic worldview were fairly reasonable. Models of systems guided by these assumptions used to yield pretty useful predictions and formed the basis of effective governance. But in a world on the brink of Singularity, as things get wilder, models based on a deterministic approach become increasingly more difficult to construct and their predictive powers diminish significantly. What exactly are the causes to the failure of models on the brink of Singularity? There are three major ones:

> **Failure of simplification methods** - Models are always simplifications of the reality they represent. The art in modeling is making these simplifications useful approximations of reality. Stable linear systems usually have a few variables of major significance and many other variables of much less influence that can therefore be disregarded without losing much accuracy. Complex reflexive systems are difficult to simplify because many of the parameters affect each other in a circular manner and it is almost impossible to disregard anything. One is left with two bad options: constructing a complicated model that contributes very little to understanding, or constructing a simple model which is rather removed from reality. In most cases, policy makers prefer the latter (and more dangerous) because it imparts some false sense of control.
>
> **Past experiences fail to produce reliable predictors of future** - In an age of accelerated change, recurrent behavioral patterns that characterize stable systems tend to disappear. Models that extrapolate future trends based on past developments will fail because the future does not resemble the past anymore. If no stable patterns can be detected, the very rationale behind predictive models is invalidated. In other words, on the brink of Singularity we can expect a serious decline in the power of models to make useful predictions. Nevertheless, current social institutions, organizations and communities base their policies on the underlying assumption that the world is a stable system, the future will resemble the past and therefore models are reliable predictors. These social bodies will become progressively more vulnerable to those unpredictable events that cannot be captured by their models.

---

[5] Brenda Hefti, a post to MadSci Network, March 15, 2002. http://www.madsci.org/posts/archives/2002-03/1016223301.Cb.r.html

**Failure of statistical models** - This is perhaps the least intuitive but most important factor. Statistical methods are used to model large populations of agents, events, interactions and so on, based on measuring a small representative sample of the population of interest. That a small sample can reliably represent a very large population is a fortunate mathematical property associated with how certain characteristics are distributed within a population. Well behaved distributions (usually associated with well-behaved systems) can produce highly reliable statistical models. If we wanted, for example, to estimate the height (or any other similar property) of an average human we do not need to measure every single human being. We can get a reliable estimate by measuring the average height of 1000 people and then know for sure that even the tallest person on earth is not very far from this average. This fact makes statistics an immensely useful modeling tool (think of insurance). Complex systems, especially human created systems and organizations, often acquire properties which are called *wildly distributed* [24]. For such property distribution there is no representative sample size large enough to help us find something useful about the whole population. For wild distributions, models based on statistical methods will simply fail. For example, contrary to height which is a biologically dependent parameter, the wealth or income of a person depends on the sociotechnological system. The distribution of wealth over the population of humans is a wild distribution also called a power law distribution. No representative sample of wealth, no matter how large, can help us estimate how wealthy the wealthiest man is. Theoretically, a single person can be richer than all the rest put together, but this cannot happen with height or weight. Wildly distributed populations are truly individualistic: the characteristics of the majority tell very little about the characteristics of individuals. Many important parameters of the sociotechnological system are wildly distributed. For example, one's wealth, one's influence in a given social network, the number of people a specific news item will reach, and so on.

As long as we are imprisoned within a view of our world as a deterministic well behaved machine, as long as we believe that modeling always works and that we can have absolute control, our systems will fail in places we least expect. As models fail more frequently they will result in ineffective governance and increasing systemic fragility. In the following sections we describe the much needed change of perspective if we are to dodge the existential risk described here. We show how the vision of a World of Views may help us to harness the same mechanisms that amplify fragility, in order to amplify resilience instead - to turn a weakness into an advantage.

## 7 Our antifragile future

If we accept the premise that the impact of technology is amplification in the broadest sense of the word, we can expect the world on the brink of Singularity to become more connected, more reflexive and more accelerating. Following the understanding developed in section 6, such a world will also become increasingly fragile. From the vantage point of our present situation, it seems that the future of the sociotechnological system will develop along one of three paths. (1) The first path leads to systemic collapse via a cascade of disastrous global events such as an economic meltdown, escalating local conflicts, collapsing of governance systems, etc. (2) The second alternative is a forceful, probably not less violent, attempt to bring progress under control (giving up much of the immense benefits it brings). (3) The third, brighter path, leads to systemic abundance and flourishing, expansion of intelligence and opening for humanity as yet unimaginable horizons of growth and transformation.

How can we modify our present so we eventually fall on the favorable path? In this section we explain why a World of Views presents a viable response scenario that overcomes the fragility inherent in our sociotechnological system. We base our reasoning on the concept of antifragility developed by Nassim Taleb [32].

### 7.1 Understanding antifragility

Our discussion of fragility in complex systems concluded that we need to focus not on specific vulnerabilities but rather on the dynamics that amplify fragility. Following this line of thought, we find little sense in trying to achieve resilience by identifying fragile points in the system and redesign them to be stronger. In a complex reflexive world we will never manage to anticipate all the fragile points which may trigger a global failure due to an unlikely accident. We need therefore a radically different approach; we need to come up with dynamics where local failures stimulate a global strengthening.

In his book [32], Taleb is trying to investigate what would such dynamics be like. He starts by asking a brilliant question: if fragile things seem to characteristically dislike volatility, disorder, uncertainty, mishappenings and stressors of all kinds, what would be the character of those things that are the exact opposite, things that *like* volatility in all its forms? He calls such things *antifragile*. Resilience does not seem the proper designation for them. Resilient things would be at best indifferent to volatility. Antifragile things are not just resilient or indifferent to volatility, they *positively love* volatility, disorder, accidents, uncertainty, stressors, chaos and so on. They love all these because they *thrive in their presence*. When it comes to a system as dynamic and complex as the sociotechnological system, it is not enough for it to be indifferent to the unavoidable volatility that is inherent in its very dynamics. The system needs to thrive on volatility - it needs to become antifragile.

What stands behind the idea of antifragility and how do we make a system gain from all those situations which intuitively seem to be detrimental? According to the paradigm of determinism and control that is deeply rooted in the Newtonian view, anything which is unpredictable, unexpected, accidental, or otherwise outside the prescribed norms, is by definition a potential risk to the system. However, unpredicted events can also be positive. If we go back to amplification as a general property of sociotechnological dynamics, ideally we would like to have some kind of selective amplification mechanism with magical sensitivity: amplifying events that can become globally advantageous while suppressing those with potential adverse effects. A system thus constructed would clearly thrive on the unexpected. Unfortunately, such a mechanism seems to require perfect predictability which is, of course, an unachievable idealization.

Yet, antifragility is not just an abstract idea. If we think about biological evolution as a kind of a technology and of life forms as its artifacts, we can clearly observe that life in general is antifragile. In the long history of life, failure is the rule and success is the exception. At every given point in time, there are many more extinct species than living ones. However, no doubt, life is thriving in spite of (or perhaps thanks to) all misfortune and this is a rather consistent trend for a few of billions of years already. What is it that makes life antifragile?

Natural evolution is the greatest entrepreneur; it achieves antifragility via endless experimentation and creation of ever more options. It keeps whatever works and scraps everything that does not, freeing up resources for further experimentation. It creates options by variation through mutation and lets the environment select which variations are suppressed and which are amplified through procreation. The most significant point is that while evolution easily disregards its failures, the successful experiments that get to proliferate naturally become the sources of new options. This is how evolution achieves its selective amplification. For example: certain cellular mechanisms (e.g. energy production, DNA replication etc.) discovered by evolution billions of years ago, are shared by almost all contemporary life forms. Such mechanisms will continue to be retained even if the vast majority of life forms at some point in time will go extinct.

From the perspective of all life, any single trial and error experiment costs almost nothing in terms of investment. Therefore, the penalty involved in a failed species (or a failed individual) is very low, just a negligible fluctuation in the overall distribution of the biological mass. But the gain involved in one successful species is immense: finding a new way to proliferate. In the long run, any such success will be shared by many life forms because the more successful a species is, the more it will originate new variants of itself.

In short, life's secret is a very smart investment strategy: it diversifies its investments as much as possible so any single failure has a very small and insignificant impact while a single success has a huge gain. The gain of one success exceeds many times over the loss involved in many failures. The net effect of such strategy is that there are many more failures than successes, but gains accumulated from successes are much greater than losses accumulated from failures[6]. Evolution does not try to predict the outcome of its experiments. Just experimenting and retaining the successful (in terms of fitness) experiments seems to be more profitable than any attempt to predict what would be fit based on past experiments[7].

Random mutations? A serendipitous speciation event? An abrupt change of environmental circumstances? A cataclysmic event of this or that kind? For life as an investor and for evolution as its investment strategy, these are all opportunities realized either as new experiments (think of species as start ups), or as stress tests for existing experiments (think of ecologies as competitive markets). As long as the gain from success is much greater than the penalty of a failure, life thrives on unexpected changes - it is antifragile.

It is important to note that antifragility is related in this context to the biosphere as a nested hierarchical structure of systems each of which is antifragile on its own and operating a similar evolutionary dynamics. For example: species are the experiments of large ecosystems, likewise, individuals are the experiments of species. The antifragility of each systemic level derives from the diversity and local fragility of the elements populating the levels below. The selective stressors are not provided only by the external environment. Variations on the same successful experiment compete with each other over resources, so selective pressures may arise from within the evolutionary process itself, driving its dynamics even while the rest of the environment is less demanding.

We believe that this evolutionary investment strategy presents a critical key in achieving an antifragile future of the sociotechnological system. To follow

---

[6] The strategy is technically described by Taleb as a convex pay function [32]
[7] This kind of strategy seems to work best under frequent, unpredictable variations in the environment while in relatively stable environments, predictive models based on past experience will perform better.

life's lessons will demand a paradigm shift in our approach to sociotechnological dynamics.

## 7.2 A paradigm shift - from control to experimentation

How do we apply these profound lessons from biology? We need first to accept that our world is becoming more complex, less predictable and therefore less controllable. As we progress, our artificial systems and environments become more lifelike: interconnected, messy and full of surprises. In response, we need to move away from the Newtonian view of the world as a deterministic clockwork to a much more organic and diverse paradigm. This is what a World of Views is all about. Systems will fare much better if they are left alone to self-organize rather than be externally organized to fit premeditated idealizations. In thinking about the future sociotechnological system we adopt life's investment strategy: creating a system which is favoring and supportive of experimentation and diversification, which always produces more options. It is not that we can know *a priori* which options will succeed but we need to eradicate our collective fear of failure and learn to allow more risks, albeit calculated ones. Take for example space exploration. With the appropriate supportive ecology, multiple parallel projects make the prospects of space travel closer compared to the undertaking of a single, state managed project. This is true even if most of the multiple projects are doomed to fail.

A paradigm shift from values of prediction and control to fast prototyping, and trial and error implies a respective shift of emphasis from modeling to hands-on tinkering of system dynamics. As models become less trustworthy in a fast changing environment, undue reliance on them exposes systems to increased fragility. It does not mean that we should stop developing models. Modeling is a very powerful and successful cognitive tool whenever the general circumstances are stable. But we need to become acutely aware of the inherent shortcomings of models especially with regard to complex, volatile systems. In this sense, the sociotechnological system is getting to a point where models and the kind of control they promise are not sufficient. We need to develop the knowledge of what makes systems thrive on volatility because such systems are going to fare best in our future.

We cannot invent, or premeditate the specific characteristics of a human or a post-human society, or how it will take shape in the volatile environment even of the very near future. We cannot predict which social institutions will be needed to sustain social order in such circumstances. But we can conceive and build a sociotechnological system which is able to self-organize, adapt and grow when exposed to a volatile future. Moreover, we will not be able to totally avoid serious accidents. Becoming antifragile is neither about prediction nor about avoidance of failures; it is about smart risk taking, experimenting, and replacing control with self-organization wherever possible. These will be the guidelines of a *distributed governance* system further discussed in section 9.

**7.3 Antifragility and a World of Views**

Human civilization as well as global social institutions, cities and communities are all complex organisms. Complex organisms cannot be designed, they *evolve*. At best we can learn how to influence and guide their evolution. In the previous section we applied evolutionary principles to the sociotechnological system. The key concept is antifragility, the property of a system that thrives on volatility. We argue that a World of Views as a future organizing framework of human civilization, social organizations, cities and communities is a vision of an antifragile future.

In a World of Views, every worldview represents a unique and integrated cognitive structure, held collectively by a network of individual agents. These cognitive structures constitute different ways of representing the world and self. Based on different biases, values and premises, worldviews in the sociotechnological system embody options to perceive and operate in the world analogous to life forms, species and ecologies. We see a World of Views as a nested, modular self-organizing structure, where worldviews occupy the highest level but in themselves are modular and diverse [30]. Diversity, modularity and openness are the essential properties of worldviews that together characterize an adaptive structure capable of containing failures while propagating successes within the system, thus realizing antifragility.

> **Diversity** - Entrepreneurship and experimentation at all scales and areas of activity will be the source of increasing diversity. In simple terms diversity means that systems and components have multiple versions, that there are always multiple ways to achieve a goal and every item or component can be used in multiple ways for multiple functions. Diversity leads to redundancy which is characteristic to organic systems and is missing from clockwork systems, designed to maximize efficiency at the expense of resilience.
> 
> The principle enabling diversity in the sociotechnological system is *converting wealth to freedom*, i.e. investing any surplus (wealth) in creating more options of choice (freedom). Similar to natural evolution, every experiment draws from the surplus of an already realized success and diverges from it to explore further options. Presently, experimentation is mostly driven by various adaptive stressors and constraints. But in the future, technology will eliminate most of them. In an age of abundance, experimentation will be driven and shaped by a universally cultivated spirit of exploration that will emerge as a new evolutionary force (see subsection 8.1).

A diversity of views will form consensual enclosures and localities even in a hyper-connected world. Agents sharing a worldview will naturally have more traffic going on among themselves while agents holding different worldviews will keep their interactions more in check. In such an environment, failed experiments will remain relatively local, while successes will multiply their gains when distributed across the membranes that keep communities sharing different worldviews apart. Yet, an ecology encouraging openness and sharing will keep everyone looking for new options that were already tried out by others, because these represent gain without risk. Such ecology will drive both competition and exchange.

**Modularity** - is an essential characteristic of worldviews in our vision. The idea is that worldviews will generally have a nested modular structure (i.e. each module is also modular), where each module is a complex of concepts, perceptions, values etc. that form together a coherent and relatively independent whole. Communities holding modular worldviews will be able to acquire, discard, experiment and share individual modules independently of other modules. The importance of modularity is in the evolutionary dynamics that it allows: 1) in modular systems, component modules can evolve relatively independently which supports the localization of failed experiments; 2) systems can exchange and share component modules which supports amplification of successes; 3) diversity of modules allows for increased flexibility, adaptability and redundancy. In analogy to vertical inheritance in natural evolution, successful modules will proliferate and be acquired by many communities, while producing myriad variations of themselves (speciation). Communities sharing or exchanging modules to augment their larger worldview will in fact mimic mechanisms of co-evolution such as cooperation, symbiosis and horizontal trait transfer[8].

**Openness** - reflects the tolerance of a worldview to changes in its own structure, its tolerance towards worldviews different from it and its capacity to constructively interact and share with other worldviews. Openness, to use a biological metaphor, is the degree of permeability of a worldview's membranes. To be open, however, does not mean that anything goes. Every worldview embodies a selection mechanism that assures its own consistency. Openness, therefore is to be understood not as an unconditional acceptance of any change but rather as tolerance and responsiveness towards change. Openness is essential in our vision because it is strongly correlated to the dynamics of experimentation and exchange without which the system cannot achieve antifragility. Worldviews that are not open towards the flow of ideas from other worldviews can perhaps develop resilience but cannot become antifragile because they resist volatility. Additionally they will be less helpful in producing options and exchange. Open worldviews not only contribute to the antifragility of the whole sociotechnological system but can themselves achieve a higher degree of antifragility.

---

[8] Horizontal exchange of traits and horizontal gene transfer is very rare in complex animals but is common in bacteria, fungi and plants.

In conclusion, adopting a perspective that sees the sociotechnological system as a complex organism, we propose here an informed speculation that if this sociotechnological organism will not collapse prematurely, it will converge towards an antifragile configuration, because this is the only configuration that can effectively cope with the volatility induced by accelerating change. We do not know what might be the particular properties of such a world but we have drawn here those properties that seem to us essential to its continuation. The processes driving towards a World of Views cannot be systematically designed. These are rather reflexive processes of self-organization, the kind of which we can already see in various social phenomena such as open source communities[9], makers movements[10], occupy movements[11], crowd sourced projects[12] and more. These example processes indeed seem to demonstrate high levels of diversity, modularity and openness.

## 8 Abundance and sociotechnological evolution

One of the greatest promises of technological progress is the promise of practically unlimited affluence; food, energy, knowledge, health care and physical safety will become ubiquitous and virtually unconstrained. Baseline standard of living and wealth for all human beings and possibly other sentient agents, will rise to what today is enjoyed only by the richest and highly privileged. There will be no limits to growth and potential self-fulfillment [9, 10]. The achievement of abundance is of course not guaranteed, yet we believe that by adopting the paradigm we develop here, sociotechnological evolution will converge towards such a state of affairs. In this section we examine the effects of abundance. We argue that besides the fact that abundance will catalyze the emergence of a World of Views, it will also profoundly change the very nature of the evolutionary process. This has to do with the freedom it will enable.

---

[9] http://en.wikipedia.org/wiki/Open_source
[10] http://www.forbes.com/sites/tjmccue/2011/10/26/moving-the-economy-the-future-of-the-maker-movement/
[11] http://en.wikipedia.org/wiki/Occupy_movement
[12] http://en.wikipedia.org/wiki/Crowdsourcing

According to the Universal Selection Theory, [16, 11, 5] an evolutionary process is a combination of three different component processes: 1) a mechanism of variation of behavior; 2) consistent selective processes; and 3) a mechanism for preserving and propagating the selected variations. For example, mechanisms of variation are realized by entrepreneurship and experimentation. Selective processes are realized by the practical application of worldviews in actual situations, while mechanisms for preservation and propagation are realized by the sharing and exchange ecology cultivated in a World of Views.

The specific characteristics of these components define the context and the nature of any evolutionary process. Yet, as it is clear from the name of the theory, the aspect of selection is predominant because selective processes are the ones that set the direction of evolution. Our sociotechnological system exists within a vast space of possible states. When it moves from state to state it forms paths that represent evolutionary and developmental processes. Paths are shaped by the selectors influencing the sociotechnological system and driving it from its current state into future states.

**8.1 Shift in the sets of selectors**

Technological progress can be viewed as a progressive elimination of environmental constraints. Transportation networks, for example, relax or eliminate the constraint of physical distances and make people considerably more free in this respect. As long as there are physical constraints, they will act as selective pressures guiding technological evolution. For example, a lack of a clean renewable supply of energy places a pressure that stimulates the sociotechnological system to innovate in that direction. Once there will be enough energy to fulfill all needs with excess, such pressure will be greatly relaxed and might even disappear. Abundance can be understood as a situation where certain sets of constraints driven by survival needs have been greatly relaxed or entirely eliminated. When constraints that guide the trajectories of the evolutionary process disappear, there arise in the system's state space, what we may call choice zones. Within such zones, various trajectories are *no longer differentiated* in relation to the current state of the system and its environment. In other words, the agent or system can choose whatever trajectory, without a discernible effect on their immediate fitness (e.g. what kind of cheese would you buy in the supermarket). Such choices, being equal against any relevant selective criteria, are clearly not guided by values imposed by survival needs.

Choice zones therefore are fields of relatively unconstrained exploration where individual agents are free. Nevertheless, agents will make different choices and form disparate paths to express their unique views and way of life. Initially, such choices are useless from an evolutionary perspective, but they do take place within an interactive and reflexive system. In the course of social interactions, these choices may gain 'evolutionary usefulness' by influencing the agent's other preferences, values, social status etc.

On the scale of the sociotechnological system, choice zones are where entirely new selectors may arise and consolidate via the interactions among agents. Such selectors will start in turn to influence the direction of the sociotechnological evolution, gradually replacing the physical adaptive pressures that used to guide biological evolution. Even in the absence of explicit physical constraints imposed by competition over limited resources, the emergence of new selectors in choice zones will catalyze a novel evolutionary motion, very different from the one driven by survival needs[13]. In other words, choice zones are cradles of new value systems.

A good example of an emergence of new selectors is the world of fashion pertaining to clothing. Fashion emerged along human history as clothing gradually started to mean more than just the protection of the body. Once the basic function of clothes was achieved by diverse technological means (fabrics, methods of manufacturing, stitching, etc.), clothing gradually became a choice zone. The fitness of agents to their environment (i.e. the probability of their survival) was no longer differentiated by what they wear. However, becoming a choice zone did not bring clothing to an evolutionary dead end; on the contrary, humans started to use clothing to distinguish themselves in other contexts. Clothing became a social medium and a complex system of signals used to express the wearer's character, style, social standing and affiliation, occupation, values and more. The physical constraints (protection against the elements) that guided the evolution of clothes as a technology, were replaced along history with other sets of selectors - means of expression on the social plane.

As long as agents are constrained by considerations of fitness, they usually operate in a reactive manner (see section 5) being forced to make certain choices in order to survive. But wherever abundance creates choice zones, agents will start to operate actively; they will express their freedom by creatively inventing new values and making meaningful selections thus distinguishing themselves in their social interactions. We can call such inventions expressions of style, aesthetic choices or simply the expression of freedom. Once such expressions consolidate they become new sets of selectors, guiding the actions of agents and eventually the direction of sociotechnological evolution.

---

[13] In a recent paper titled Evolvability Is Inevitable: Increasing Evolvability without the Pressure to Adapt [22], the authors bring experimental evidence that the evolvability of a population of agents (i.e. their capacity to produce and inherit significant phenotypic variations) can grow even in the absence of adaptive pressures. This research supports our hypothesis that selectors emerging in zones free from adaptive pressures can eventually have significant evolutionary effects on the sociotechnological system.

The conditions of abundance will bring with them a sweeping shift in the selective forces that guide the evolution of the sociotechnological system. The shift will be from survival driven selectors (reactive forces) to selectors shaped by the need of agents to give a meaningful expression to their freedom (active forces). Agents existing in conditions of abundance and for whom the concerns of survival and replication have become redundant and almost entirely taken for granted, will actively and creatively seek to differentiate themselves in entirely new ways.

The novel selective forces that will emerge, will derive from worldviews adopted by agents. The expression of these forces will be primarily aesthetically oriented, i.e. shaping the environment to reflect style self-description and self-actualization, rather than be driven by survival needs. They will constitute, we believe, the dynamic medium of society's next evolutionary phase - a World of Views.

## 8.2 A World of Views as the active expression of freedom

The emergence of new sets of selectors is not entirely unfamiliar. Already today, at least in some parts of the world, many of the choices that we make as individuals are rather expressions of freedom than adaptations to constraints: e.g. choice of profession, style of living, social group, faith, etc. What is going to be significantly disruptive in the future is the immense prominence of such choices in our individual lives and the overall social dynamics. We will become true creators and makers, and this is one of the more profound outcomes of intelligence escaping its biological constraints. A World of Views, the kind of future that will be an expression of our ever growing freedom, is also an antifragile abundant future.

Shaping our future as an active expression of freedom and not as a response to existential stressors will require a very profound change in humanity's most fundamental assumptions about existence and according behaviors. Already John Maynard Keynes [20] anticipated the profound disorientation and loss of meaning that might occur when a society achieved a condition of abundance but continued to deal with it as if there was continuing scarcity:

> "The economic problem, the struggle for subsistence, always has been hitherto the primary, most pressing problem of the human race [...] Thus we have been expressly evolved by nature with all our impulses and deepest instincts for the purpose of solving the economic problem. If the economic problem is solved, humankind will be deprived of its traditional purpose. Thus for the first time since his creation man will be faced with his real, his permanent problem, how to use his freedom from pressing economic cares [...] There is no country and no people, I think, who can look forward to the age of leisure and of abundance without a dread. For we have been trained too long to strive and not to enjoy."

We do not understate the inertia of human collective and individual psyche and history, which is a very powerful force in shaping sociotechnological development leading to the future state of the world. But our vision is one of overcoming the reactive nature of human intelligence and transforming it into an active expression of freedom.

If we succeed, aesthetic choices, philosophical commitments and elegant belief systems will become the shapers of our existence, much more than they are today. Throughout human history, people have been modifying their appearance, their behavior and their environment to suit and express their worldviews. This is how cultures were shaped. Whenever feasible, even the human body has been used as a medium of expression with methods ranging from cosmetic body modifications, to chemical modifications of metabolism and behavior (e.g. consuming mood altering substances such as coffee, sugar, alcohol, cannabis, etc.). A relatively rare and more extreme example is that of gender transformation which involves complicated surgery procedures, prolonged hormonal treatments and profound behavioral and psychological adaptations.

In the coming future, we can envision the gradual emergence of technologies for both body and brain modifications including genetic modifications, nanotechnological implants and more (e.g. changing one's skin color, muscle build, sensory abilities and much more). Brain modification technologies are of special interest because of their impact on the social fabric. For example, procedures such as memory editing, real-time filtering of experiences and personality modifications (e.g. turning an introvert into an extravert) are only a few of the options that neuroscience promises[14]. Human agents may radically modify their perception of physical reality using brain-machine interfaces or genetic manipulations. Individuals may merge their minds to form multi-bodied collectives or extend their minds to other bodies, biological or otherwise. For example, people who view reason as the highest value in existence may modify the way their brain operates to accentuate the expression of this value in their lives. They will be able to literally become living versions of Star Trek's Vulcans.

---

[14] Authors@Google: Ramez Naam, Nexus. http://www.youtube.com/watch?v=dlHAFHOsp9Q

Such developments will enable massive exploration and diversification into mind species or intelligence species with very different perceptions of reality. When diversity of views will deeply transform the very world we live in and will become the primary selective force that guides evolution, the sociotechnological system may undergo a Cambrian explosion of co-evolving kinds of intelligence.

**8.3 Bootstrapping a World of Views**

A World of Views is presented here both as a key strategy to achieve an antifragile existence under conditions of accelerating sociotechnological change, and as a plausible outcome of abundance. The achievement of a state of abundance, in turn, relies on avoiding the obvious predicaments of a highly volatile accelerating dynamics, without trying to stop it. The interdependence among the three factors of diversity, antifragility and abundance, constitute a positive feedback operating at the highest scale of the sociotechnological dynamics. It is the mutually reinforcing influences of its components that will bootstrap a World of Views.

It is important to note that for the purpose of bootstrapping it will be enough to achieve such levels of abundance so that aggressive competition will not make economic or political sense anymore. Still, abundance is not only a technological issue, nor is it an economic or political one. Abundance is also a psychological state of individuals and the collective human mind. It is a state where survival instincts, fear and aggression conditioned by eons of natural evolution are replaced by novel forces of motivation. We must not forget that a profound psychological transformation is a necessary aspect of sociotechnological evolution. After all, human minds and the way they process information and react are primary components of the sociotechnological system.

What is the concrete organization that would be capable to facilitate this bootstrapping process and bear the diversity of a World of Views? Our vision requires a common platform that serves both as a universal medium of interaction and exchange, as well as a maintenance facility that safeguards operational integrity and the conditions of abundance. This will be the function of *a distributed social governance system* which is our next and final topic of discussion.

**Part III Paving a multiway to the future**

**9 Distributed social governance**

Bootstrapping a World of Views requires a governance system which facilitates three major functions: 1) providing the necessary sustainable platform for a future society; 2) allowing for the co-existence of diverse intelligences and their embodiments in a shared physical and computational space and 3) providing a medium for evolution of intelligence through communication, dialogue and co-evolution among diverse intelligent agents.

**9.1 Away from sustaining order**

Current social organization is based on the deeply ingrained cultural belief in the inevitability of hierarchical governance based on authority and enforcement. This belief is obviously married to the clockwork view of the world (see 6.3) as a mechanism composed of bolts, nuts, gears and screws which can be individually identified, cataloged and registered for later usage[15]. Nuts, gears and screws, in this case, are human individuals and in order for society (the clock) to function, they have to be positioned within the social system exactly according to their specifications. Moreover, misbehaving components should be repaired or replaced like broken parts of a machinery. While in some progressive segments of civil society, this worldview is already considered outdated, not so is the case on the sociopolitical plane, where the world is broken into regions and nation states with governments comprised of functionally separated ministries, which in turn have separate departments, etc. The whole organization of governance distinctly has a top-down control structure. At the very bottom of this hierarchy, there is the citizen who even with all the virtues of a democratic political system available only in some places, is pretty much bound by a rigid system of formal rules and behavioral patterns (e.g. basic education, taxation, military service, religion, cultural behavior etc.). We are not going to delve into political philosophy here, but we assert that the hierarchical governance paradigm based on the mechanistic clockwork metaphor is absolutely incompatible with the requirements of the future social organization we envision. No matter how open and allowing are the operational attitudes of such a hierarchical order, it will resist the free explorative dynamics of a World of Views, due to its inherent structural principles.

There is even more important functional reason why hierarchical governance systems fail as the world becomes increasingly hyperconnected and complex. The prevailing bureaucratic mode of operation of organizations at all levels of society [25] assumes predictability and slow change in the world. These assumptions are largely obsolete today, let alone in a world approaching the brink of Singularity. Such *modus operandi* will become less and less effective with the current trend of sociotechnological development. It will increase the fragility of the system rather than ensure its stability and resilience. In a situation of accelerating change, hyperconnectivity and reflexivity, limiting the variability of responsive behaviors of the components of a system, lead to increased fragility (6.2). In hierarchical organizations, a shock that causes the top levels to fail may paralyze the whole organization [15].

---

[15] In social and management context this view was developed at the end of 19th century in the form of a scientific management paradigm by Frederick Taylor [33].

As we already mentioned above, what is largely overlooked by the current sociopolitical paradigm is that global society, by its very nature, resembles a living organism much more than an artifact designed by humans. As a complex organism, it evolves and self-organizes rather than be externally controlled (see 7.3). The vision of a World Government, an institution responsible for regulating the global affairs (e.g. regional conflicts, poverty and inequality, environmental degradation etc.) of a hyperconnected world [23, 6] is just another control layer on top of existing hierarchical structures. As such it is infeasible, given the accelerating and unpredictable nature of sociotechnological evolution. A self-organizing form of social governance based on dialogue and mutual agreement at multiple levels [6] and not on idealized top-down control structures, seems to be the only alternative to anarchy and dystopia on the brink of Singularity.

## 9.2 Towards organizing disorder

In a World of Views we envision a paradigm shift in governance, from maintaining a top-down prescribed order to spontaneously organizing bottom-up disorder, thus allowing the social organization to embrace volatility and uncertainty which are salient features of a future society. We propose distributed social governance as a medium that enables the self-organization and evolution of social institutions and structures. In the distributed social governance paradigm, the rules of the game are not about setting values and operational norms (e.g. protecting the weak from the powerful, ensuring free markets, or preventing environmental degradation). Rather, they are about coordinating the interactions among the agents in the social system, in order to achieve a sharing of meaning and values as well as mutually beneficial ways of co-existence, co-evolution and cooperation towards collective goals. There is no dictation of an *a priori* given value system on the agents that constitute the social fabric. In this sense, a distributed social governance system is not the embodiment of the governance regime itself, but rather "an exceedingly complex array of actors and institutions in forms of distributed and disaggregated governance that exist in a shared conductive medium [38].

Within such a medium, intelligent agents would operate according to their own ethical, aesthetic, professional and technical norms derived from worldviews which are emergent and not imposed externally. Assemblages (coalitions) of agents will self-organize into *ad hoc* management systems or institutional structures according to the needs and views shared among participating members leading to a system of multiple overlapping structures, organizations and institutions [7, 37]. Such form of distributed governance will be antifragile. Disintegration of any single organization will have almost no impact on the whole system, because other overlapping organizations will adapt to the new situation and take over the functions of the failed one. In a World of Views, governance is no longer synonymous with control and stability, but rather a provider of well-structured spaces of engagement that facilitate the communication and exchange among diverse actors without compromising the choice of multiple options nor the effective criteria for consensual selection [38].

It is plausible to assume that in a World of Views with distributed social governance there will be no nation states as sovereign territorial units. But it doesn't mean that these will be replaced by some kind world citizenship. Social groups will be taking the place of nation states. Such groups will be much less identified by elements such as physical territory, economic status, national identity or language. They will assemble themselves according to the worldviews shared by their participating members. Worldviews will serve as the primary coalescing element of an otherwise fluid collective identity. They will form kinds of trust realms or virtual states [29] each embodying a collective view. In a World of Views, joining or disengaging from a view will be straight forward. An agent will habitually join more than one view. This freedom and multiplicity of affiliation will encourage the formation of fluid and overlapping social contours. Most of the social institutions will not be based on territory or the distribution of other physical resources. Governments will not distribute tax money to various projects, but instead, resources will be distributed according to the explicit wills of the individual participants of each virtual state or trust domain. If an individual thinks that sending astronauts to Mars is a worthwhile endeavor, she, he or it may initiate or join the project by investing whatever financial, intellectual or other resources, available to them.

Institutions under such governance paradigm will have the capability to transform themselves in a much more profound way than the traditional institutions of nationally organized political democracy. They will lack the formal legitimacy of representative democratic institutions, and have constituencies which are fluid and boundaryless [38]. These institutions will have no persistent identity of their own, eliminating their tendency towards self-preservation after their actual societal functions have long been exhausted[16].

The way to create such a distributed social governance system is by following principles consistent with antifragility: 1) fast and cost effective prototyping of new social structures as well as a swift discarding of failed ones; 2) diversification of options; 3) iterative self-organization in response to the impact of unpredictable events and 4) maximal capitalization on success. Conventional wisdom implies that such a 'non-systemic' system would be a mess, and probably culminate in a destructive anarchy. Yet, examples like the open source software movement show the opposite, suggesting that the paradigm described here could actually be more productive and successful than the conventional hierarchical one. Open source projects are often started as initiatives of single developers performing creative experiments, many of which never reach a useful application. However, the successful ones are sometimes substantially more advanced than software products developed by multinational corporations allocating huge amounts of resources for the task. For example, *Apache HTTP Server*, the leading web server software, serving about 51% of world's Internet servers (about 331 million, including US State Department) was started by an informal group of eight programmers in 1995. The closest competitor, *nginx*, is also an open source project started in 2002 by a single person[17]. The success of the open source software movement triggered a spontaneous adoption of the model in other fields e.g. Maker Movement[18] and Open Science[19].

---

[16] Reflexivity and Fallibility: conversation with G.Soros.
http://www.youtube.com/watch?v=BwbSKZMerhw&feature=youtube_gdata_player

**9.3 An intellectual technology of building shared realities**

*Distributed social governance is a collective and conscious use of social system design methodologies and tools comprising an intellectual technology by which alliances of intelligent agents build shared representations of reality, visions of the future and seek their realizations* [2, 34]. Diverse populations of agents utilizing such technology will produce a multiplicity of social system design processes. This multiplicity will give rise to an environment where worldview based alliances are actually *constructing shared realities* according to their visions and mutual discourse.

The heterogeneity of worldviews and existential styles of intelligent agents will inevitably give rise to a diversity of respective environments, virtual states and social institutions reflecting these worldviews and their amalgams. This diffused yet distinct multiplicity will constitute the ecology of the future society where social actors and their alliances co-evolve along multiple paths. Co-evolution can be realized only by searching and nurturing shared views among social actors and by that, building a consensual reality where communication and collaboration are effective. Yet, in a World of Views, the option of co-existence without reciprocation, will be an available option of any social actor. Co-evolution and cooperation, beyond the necessary support in the global platform of sustainability, will always be a matter of choice at all levels. This freedom is the ultimate promise of a future of abundance.

---

[17] April 2013 Web Server Survey. http://news.netcraft.com/archives/2013/04/02/april-2013-web-server-survey.html
[18] http://en.wikipedia.org/wiki/Maker_movement
[19] http://en.wikipedia.org/wiki/Open_science

The hypothesis of social constructivism that "the world is socially constructed in two related senses, as distributed cognition and as shared realities" [16] acquires a qualitatively new meaning in our vision. Social systems have a property of reflexivity, well described by the so called Thomas theorem: if persons define situations as real, they are real in their consequences [25]. Images and models of the social system held by its participants are coupled with the actual properties of the system via the actions and perceptions of agents. These cybernetic relationships are the key to understanding the evolution of the sociotechnological system. Where biological and technical constraints are progressively removed and social actors are free to shape their environment as an expression of their worldviews, the implication of reflexivity in a heterogeneous world is the absence of a single convergent path of social development. Rather, it implies interaction among bundles of diverse paths and their co-evolution without the guidance of overarching principles - *a multiway to the future*.

A distributed social governance needs of course an infrastructure, a medium for collaboration and a culture of dialogue. In the case of open source software development that we already discussed, such a medium provides individual developers and groups with a rich toolkit enabling them to contribute to each other's code in the same programming language, and introduces a set of rules of conduct facilitating higher levels of collaboration. These tools define the size and complexity level of a project realizable by a distributed effort. Likewise, distributed social governance at the scale of the sociotechnological system should allow for interactions between humans, post-humans, AI agents and other machines, facilitating the creation of shared symbols, meanings and worldviews. Given the present difficulties of communication and cooperation between much less diverse groups of humans, the enabling intellectual technology required for distributed social governance presents a difficult multidisciplinary challenge.

We see education as the single most critical catalyst of the emergence of a World of Views and distributed social governance as its enabling platform. Towards the realization of a World of Views, educational systems need to be transformed. The intellectual technology for constructing shared realities will have to be instilled and shared by all members of society of all views including radically divergent intellects such as post-humans or AI agents, if they are to become members of such future society. The arts of dialogue, negotiation and exchange in a volatile reflexive environment are the primary skills necessary for constructing such shared realities. In a highly diverse social dynamics these skills will be the unifying and integrating forces holding a World of Views together.

## 9.4 Closing the gap between reality and vision

The prevailing way of thinking about how to induce and guide change in a system is characterized by a general scheme of systemic change, comprised of three steps: (1) identifying the current situation; (2) identifying the image of a future desired situation; and (3) figuring how to reach from the current situation to the desired one, that is, closing the gap between reality and vision [13, 34]. We can identify three approaches of addressing the third step: *strategic planning*, *strategic navigation* and *strategic exploration*.

> **Strategic planning** - is aimed at developing a procedure, as detailed as possible, for achieving goals [13]. For example, in the case of sea voyage, it is plotting a course on the map from the home port to the destination port and then following it closely throughout the entire journey.
>
> **Strategic navigation** - takes into account unpredictable circumstances along the way and allows much more freedom for adjusting the ship's course and speed depending, for example, on weather conditions along the route or maybe icebergs that might block it [14].
>
> **Strategic exploration** - When the geographic position of their destination was largely unknown, and with only partly reliable navigation methods, the great explorers of the so called Age of Exploration determined the course of their ships based on guess work and hearsay that often proved fatal. Nevertheless, since there were a lot of expeditions, new lands were discovered and more or less reliable routes were eventually established. If a large and well equipped ship finds itself in the middle of the ocean without a map or any other way to determine its whereabouts, the most effective method to find land would be to launch a number of small boats in different directions. The combined probability of one of the boats to find land is much greater than if the ship was trying to figure the most probable direction and follow it. This method comes of course with a risk, as many of the boats may not return at all.

The importance of planning, navigation or exploration in bridging the gap between reality and vision depends on how reliably we can predict the future based on our past experiences. In the case of the sociotechnological system, strategic exploration becomes more effective when our past experiences fail to inform us about future events due to hyperconnectivity, reflexivity and accelerating change (see 6.2). We therefore propose strategic exploration as the best approach for directing the sociotechnological system towards a World of Views. Strategic exploration implies the distribution of most of the system's resources among iterative moves towards the desired direction without extending them beyond the planning horizon. Once the planning horizon is reached, a new current situation needs to be identified and the gap to be bridged, reiterating the same procedure.

In order to bridge the gap between the current state of affairs and a World of Views, what seems to be within our planning horizon is a careful dissolution of the hubs of power and cultivation of much more fluid hierarchical structures. Practically, this amounts to promoting, facilitating and developing trends which are already in motion:

> **The rise of alternative currencies** A purely peer-to-peer version of electronic cash allows financial transactions to take place without the mediation of a trusted financial institution[27]. Cryptocurrencies (e.g. Bitcoin[20]) and other alternative currency systems (e.g. Flowplace[21]) operate without central authority and the hierarchical banking system that currently controls most financial transactions. They allow both issuance of currency and transactions to be carried out collectively within a network of peers without the reliance on trusted third parties. Alternative currency systems, therefore, promote the gradual dissolution of present financial power hubs.
>
> **The rise of collective decision systems** The technology of Internet and social networks offer a feasible way of exercising direct democracy without relying on mediating hierarchical structures (state institutions) to support it [28]. For additional examples see: LiquidFeedback[22] and MorsiMeter![23].
>
> **The rise of Internet activism** On-line activist networks (e.g. Avaaz[24]), probably having their roots in quite conventional networks of non-governmental organizations, use Internet technologies to facilitate self-organization of interest groups capable of exerting considerable influence on the political establishment thus disrupting its rigid power structures.
>
> **Changing the topology of sociotechnological networks**
> Sociotechnological networks are those networks that organize the flow of information and goods within the sociotechnological system. The prevailing topologies of these networks are scale-free[1] implying self-reinforcing and self-sustaining power hubs within the system. We see important new trends introducing much more plasticity into the configuration sociotechnological networks:

---

[20] http://bitcoin.org/en/
[21] http://flowplace.webnode.com/droplets/
[22] http://liquitfeedback.org/
[23] http://www.morsimeter.com/en
[24] http://avaaz.org/en/

- **Internet** has started as a truly distributed system where every node in the network was fairly equal. In time, the Internet evolved into a scale-free network structure with many unimportant regular nodes and only a few huge hubs (e.g. Facebook, Google, Amazon) controlling most of the information and its traffic. Such network topology is prone to attacks, control, surveillance and corruption[25] [1]. Peer-to-peer and distributed computing technologies[26], including multicast packet routing schemes[27], grid computing[28], fog computing[4], distributed messaging systems (e.g. Bitmessage[29]) and distributed file systems (e.g. Git[30]) contribute to the overall plasticity of the Net.
- **Edunet** is a network of collaborative production, sharing and use of the educational resources without institutional control over the development of educational programs, curricula and norms. We see the roots of Edunet in the currently emerging movement of Massive Open Online Courses[31], semantic web technologies (e.g. Google's Knowledge Graph[32]) and self-organizing knowledge networks as proposed in[17].
- **Enernet** is an emerging network of distributed energy production, sharing and storage supported by technologies like Smart Grid[33], microgeneration[34] and distributed energy storage[35]. The Enernet will enable its users to become producers and distributors of energy[36]. The emergence of the Enernet on a large enough scale will have a profound disrupting impact on the geopolitical determinants of hierarchical power structures (e.g. nation states, energy corporates) and the struggle among them.
- **Matternet** is a distributed system of design, production and delivery of manufactured material goods (hardware in the broad sense) supported by DIY and Makers movements[37]. Current technologies facilitating the emergence of the Matternet are 3D modeling and 3D printing[38] which will allow the sharing of physical designs and their production without the need for large scale factories and distribution systems.

---

[25] The NSA Files, http://www.theguardian.com/world/the-nsa-files
[26] http://en.wikipedia.org/wiki/Distributed_computing
[27] http://en.wikipedia.org/wiki/Multicast
[28] http://edutechwiki.unige.ch/en/Grid_computing
[29] https://bitmessage.org
[30] http://git-scm.com/
[31] http://en.wikipedia.org/wiki/Massive_open_online_course
[32] http://www.google.com/insidesearch/features/search/knowledge.html
[33] http://en.wikipedia.org/wiki/Smart_grid
[34] http://en.wikipedia.org/wiki/Microgeneration
[35] http://www.renewableenergyworld.com/rea/news/article/2013/07/the-case-for-distributed-energy-storage
[36] http://www.wired.co.uk/magazine/archive/2012/02/ideas-bank/energy-sharing
[37] http://www.forbes.com/sites/tjmccue/2011/10/26/moving-the-economy-the-future-of-the-maker-movement/
[38] http://en.wikipedia.org/wiki/3D_printing

## 10 Conclusion

Our envisioning of the brink of Singularity begins with redefining Singularity as an historical process, rather than an event. It is the process of continuous intelligence expansion since the beginning of human civilization. We emphasize the value and significance of the continuity of this process rather than the intermediate stages through which it passes. By that, we position the brink of Singularity situation within the continuum of human evolution, the evolution of life and evolution as a universal process. We raise the question of what would be the desired configuration and dynamics of the sociotechnological system able to facilitate open-ended intelligence expansion. A World of Views is our vision of such a configuration. We then argue why a World of Views is likely to be the only feasible configuration capable of sustaining the Singularity as a process of intelligence escaping its biological constraints and beyond. Finally, we propose distributed social governance as a bootstrapping mechanism for a World of Views and link it with the current momentum of the sociotechnological system.

At the basis of the evolutionary shift humanity is undergoing on the brink of Singularity is the progressive process of entering into symbiotic relationships with its technological artifacts. This symbiotic convergence deemphasizes the anthropocentric perspective in regard to the future. Furthermore, the past consensual understanding of what constitutes our humanity cannot serve us effectively under circumstances of accelerating sociotechnological change. From the social perspective, the most important are those artifacts that augment social interaction of intelligent agents as currently the Internet primarily is. Such artifacts do not only change us individually, they transform the very fabric of human civilization. We take therefore a systemic approach, first by focusing our discussion on the dynamics of the sociotechnological organism humanity is becoming, and second by introducing worldviews as the relevant units of evolution of sociotechnological organisms.

Our analysis of the sociotechnological evolution shows that circumstances of hyper-connectivity, reflexivity and acceleration beyond their many obvious benefits expose the sociotechnological system to fragility that will only increase in the near future and may lead to some catastrophic though yet unpredictable consequences. In order to counter this systemic effect we apply the concept of antifragility - the property of systems that thrive on volatility and uncertainty - and conclude that antifragility is necessary to secure the sociotechnological system from devastating catastrophic events. To that end, we need a paradigm shift towards what we call a World of Views. A World of Views is a nested, self-organizing structure, where worldviews occupy the highest level but in themselves are modular, open and diverse. Diversity, modularity and openness are the essential properties that together characterize an adaptive structure capable of containing failures while propagating successes within the larger system, thus realizing antifragility at multiple scales.

An antifragile sociotechnological system, however, is much more than just dodging existential risks. We argue that the benefits of technology will gradually transport humanity into an age of abundance, which will in turn have profound effects on sociotechnological evolution. This self-amplifying reciprocity will result in decline and even disappearance of evolutionary pressures that arise from limited resources and survival needs. We propose that abundance will catalyze active expressions of freedom that will become novel evolutionary selectors. In our vision of the future, the expression of freedom rather than survival is the ultimate driving force of evolution. We conclude that the World of Views is a catalyst of future abundance, which in turn reinforces the dynamics intrinsic of the World of Views. This positive feedback mechanism, once set in motion, will bootstrap the sociotechnological system towards a World of Views.

Finally, we introduce in broad lines a distributed social governance system that we foresee as instrumental to the development of a World of Views. Distributed social governance system is the implementation of a World of Views on the social plane. It is a radical extension of a democratic governance regime in a sense of abolishing the single unified paradigm, in favor of continuous construction and dismantling of experimental models that partially work. It is clear to us that global education systems are the essential key towards distributed social governance, teaching us to live in a world without survival constraints, giving up the idea of a single value system, constructing individual and shared realities and constantly innovating on them.

A transition from our current hierarchical power structures to distributed self-organized ones is of course not certain. Nation states and other contemporary organizations tend to be self-persistent and self-reinforcing (e.g. army generals inventing and initiating wars to justify the existence of armies). We believe, however, that conservative structures will either disintegrate or adapt due to their exposure to accelerating change. Those that will adapt, will eventually transform to become open and modular because such properties will characterize the best strategy of operation within the accelerating dynamics of the future sociotechnological system. The realization of a World of Views and distributed social governance system does not promise or assume a peaceful and safe world for every individual. Yet, it does safeguard the continuity of intelligence expansion and the affluence it promises.

We are aware that human nature itself could impose a serious impediment. Some claim that human nature is selfish and brutal and these traits are an inevitable consequence of our biological heritage as well as a primary shaping factor of our future. Yet, we believe that intelligence, escaping its biological constraints, will present us with the actual means to overcome this inevitability of a nature shaped by a struggle for survival. Having said this we do realize that our vision indeed requires a fundamental transformation of the human psychological construct, individual as well as collective.

We have barely tapped the tip of the iceberg of our sociotechnological future which is mostly submerged in the waters of unpredictability. In one of Star Track's episodes[39], Captain Picard, the 'natural' human, gives Data,, the 'post human', a lesson: It is possible to commit no errors and still lose. That is not a weakness. That is life. Inasmuch as one can commit no errors and still lose, one can also commit many errors and still win, or at least keep on playing. That is life. As a young civilization, still making its first steps, we choose to read in this an optimistic note: trust life and trust our wits while trying very hard not to commit those errors which are fatal. As our way into the future will require a lot of trial, we'll have to cope with many errors and their consequences. That is not a weakness. Like princess Scheherazade from 1001 nights, determined to live another day, we tell another story, the long tale of the Singularity, as long as the history of the human. Realistic as we claim it to be, admittedly with a touch of the fantastic; it is for sure an ultimately challenging vision.

---

[39] Star Trek: The Next Generation, Peak Performance", http://en.wikipedia.org/wiki/Star_Trek:_The_Next_Generation